\documentclass[aps,pre,showpacs,amssymb,a4paper]{revtex4}%
\pdfoutput=1
\usepackage{amsmath}%
\usepackage{amsfonts}%
\usepackage{amssymb}%
\usepackage{graphicx}
\usepackage[english]{babel}
\usepackage{color}

\begin{document}

\title{On the role of composition entropies in the statistical mechanics of polydisperse systems} 
\author{Fabien Paillusson, Ignacio Pagonabarraga} 
\affiliation{Departament de fisica fonamental, Universitat de
  Barcelona, Marti i Franques, 1 , 08028, Barcelona, Spain}
\email{fpaillus@gmail.com} 

\pacs{45.70.-n,45.70.Cc,05.90.+m} 
\date{\today}

\begin{abstract}
Polydisperse systems are commonly encountered when dealing with soft matter in general or any non simple fluid. Yet their treatment within the framework of statistical thermodynamics is a delicate task as the latter has been essentially devised for simple --- non fully polydisperse --- systems. In this paper, we address the issue of defining a non ambiguous combinatorial entropy for these systems. We do so by focusing on the general property of extensivity of the thermodynamic potentials and discussing a specific mixing experiment. This leads us to introduce the new concept of composition entropy for single phase systems that we do not assimilate to a mixing entropy. We then show that they do not contribute to the thermodynamics of the system at fixed composition and prescribe to substract $\ln N!$ from the free energy characterizing a system however polydisperse it can be. We then re-derive general expressions for the mixing entropy between any two polydisperse systems and interpret them in term of distances between probability distributions and show that one of these metrics relates naturally to a recent extension of Landauer's principle. We then propose limiting expressions for the mixing entropy in the case of mixing with equal proportion in the original compositions and finally address the challenging problem of chemical reactions.
   \end{abstract}

\maketitle

\section{Introduction} 

Continuous polydisperse systems are ubiquitous in Nature and in our everyday life and technological experience. Yet, their treatment with statistical mechanical tools is delicate in many respects and different approaches have been proposed to characterize them \cite{Salacuse84, Sollich98, Sollich01, Sollich02, Warren, Will, Ignacio10}. Among the existing problems, the combinatorial entropy in these systems and its role in their thermodynamical behaviour, is more conceptual. It poses a problem because, depending on the way one rationalizes the need of the $N!$ in usual statistical mechanics of identical particles, it gives rise to different prescriptions regarding the combinatorial entropy of continuous mixtures. The nowadays standard {\it bottom-up} approach to the existence of the $N!$ in statistical mechanics seems to be originally due to Uhlenbeck and Gropper \cite{Uhlenbeck32} and considers that statistical averages over microstates are done over quantum states ultimately characterizing identical --- indistinguishable --- particles, bosons or fermions, and the {\it spin-statistics theorem} ensures that the corresponding quantum states are either fully symmetric or antisymmetric \cite{Cohen-Tannoudji}. The idea consists then (although it has been recently questioned \cite{Seglar14}) in expressing partial traces over these states as full traces corrected by a simple $N!$ factor \cite{Balian, Huang}. In this interpretation two interesting consequences are worth pointing out: a) extensivity of the thermodynamic potentials arises, ultimately, as an emergent property of quantum mechanics and b) because a continuous mixture has no two particles which are exactly the same \cite{Salacuse84}, there should not be any $N!$ entering the partition functions that describe them. 

There also exists a {\it top-down} approach originally due to Gibbs and made more clear by others \cite{Sollich01, Warren,Jaynes92, Swendsen06, vanKampen84, Daan14} that consists in introducing a $N!$ in the phase integral expressions of the thermodynamic potentials so as to avoid inconsistencies with thermodynamics which statistical mechanics ought to reproduce in the thermodynamic limit. This interpretation favors an {\it epistemic} origin to the $N!$ while the quantum one favors more an {\it ontic} one. Regarding polydisperse systems, the epistemic approach does not prescribe anything in particular and different strategies have been used in the past to characterize their combinatorial entropy \cite{Salacuse84, Sollich01, Sollich02, Warren}. 

In this article, we will follow the epistemic reasoning to address the problem of the combinatorial entropy in polydisperse systems. To this end, we first recall key concepts and results about extensivity, Gibbs' thought experiment and basic statistical mechanics to introduce the notion of {\it composition entropy}. We then study more formally the problem of mixing between any two polydisperse gases and derive a general expression for the corresponding entropy variation. Focusing on the contribution to the mixing entropy coming only from the difference in composition, we find that it is a metric in the space of probability distributions and formulate an interpretation of it based on a recent extension of Landauer's principle. After having suggested some limiting expressions for this metric, we then end by discussing possible extensions of the previous analysis of mixing when it occurs together with chemical reactions.

\section{Extensivity and Gibbs' thought experiment}
Going back to the basics, a first question to be asked is {\it why should the thermodynamic potential of a system in a certain state be extensive in the first place?} The answer to that question has to do with the experimental fact that the thermodynamic state of a system has some {\it scale invariant} properties. These are the ratios of the extensive variables that characterize the state of the system or functions of those ratios: {\it intensive variables} are such functions for instance. If one were to scale up the system size by multiplying all the relevant extensive variables by a factor $\lambda$, this would leave unchanged the state of the system as described by its scale invariant state variables. In this context, requiring an extensive thermodynamic potential is equivalent to saying that it will change in a trivial manner as the system size is changed in the aforementioned way {\it i.e.} it will simply be multiplied by $\lambda$. 
\begin{figure}
\vspace{5mm}
 \hspace{-5mm} $(a)$  \hspace{36mm}   $(b)$\\
  \includegraphics[width=0.49\columnwidth]{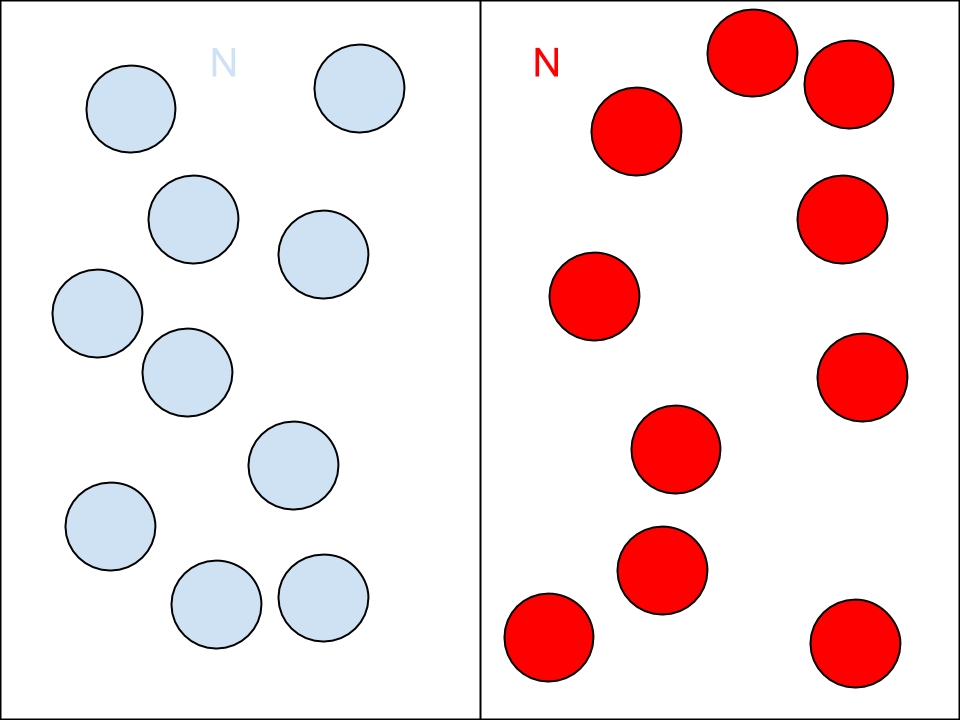} 
  \includegraphics[width=0.49\columnwidth]{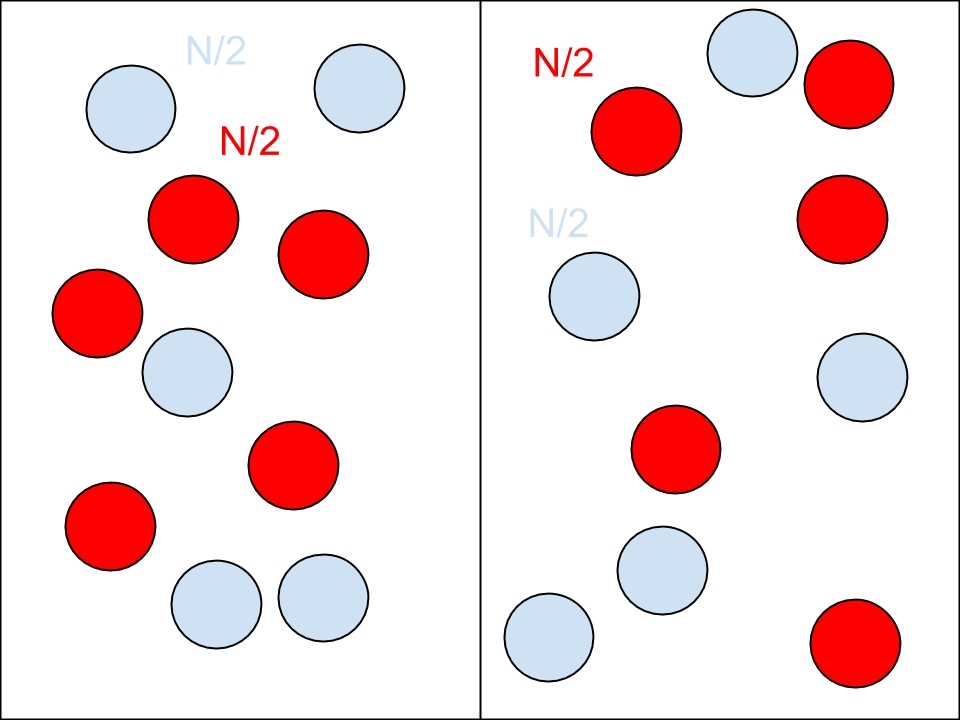}\\
 (c)\\ \includegraphics[width=0.49\columnwidth]{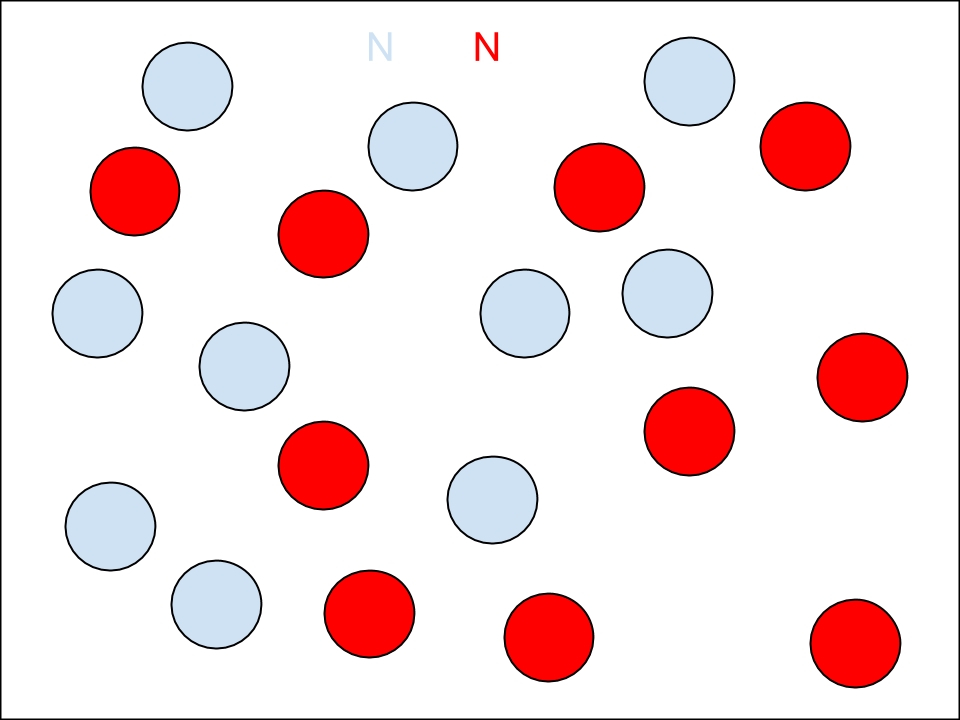}
  \caption{\label{fig1} Schematic representation of Gibbs' thought experiment. Starting from the premises that a scaling transformation doesn't change the entropy per particle implies that the mixing entropy contribution to a free energy change $\Delta S^{mix}_{b \rightarrow c} = 0$ since the transformation from $(b)$ to $(c)$ can be interpreted as a size scaling. However, the transformation from $(a)$ to $(c)$  is not a size scaling and the total entropy change needs not be zero.}
\end{figure}
In other words, the ratio of the entropy by any --- relevant--- extensive state variable is also a scale invariant thermodynamic variable. This is important because for isolated thermodynamic systems, the entropy plays a crucial role for asserting whether or not a particular transformation is irreversible. Hence, what extensivity tells us is that scaling up (or down) the system size {\it is not} an irreversible transformation.\newline The fact that scaling of the system size is not an irreversible transformation was ingeniously pointed out by Gibbs' famous thought experiment \cite{Jaynes92, vanKampen84, McQuarrie, Balian, Huang, Warren} depicted in Fig.\ref{fig1}. The idea --- here applied to a canonical ensemble --- is to compare the mixing entropy contribution $\Delta S^{mix}$ to the free energy change in a thermostated system with two compartments in two different experimental cases: 1) the mixing of two different gases --- where one witnesses an entropy increase owing to mixing --- and 2) the mixing of two identical gases --- where there is no entropy change because it is a scaling transformation ---. The cause of the so called {\it Gibbs' paradox} that arose then was that for a given ideal gas with $N$ particles, the original Boltzmann entropy $S_B = k_B \ln C V^N T^{3N/2}$ (where $C$ is a constant) was lacking a $-\ln N!$ to be consistent with both outcomes of the experiment. It is interesting to notice that the term ``\ identical ''\ for case 2) is subject to interpretation. Traditionally, it is understood that ``\ identical ''\ means that every single particle of the gases in case 2) belongs to the same chemical species. However, this seems unnecessarily constraining. One can imagine the transformation from $(b)$ to $(c)$ in Fig.\ref{fig1} for instance and it would {\it a priori} fulfill the conditions of ``\ identical ''\ gases. 
\begin{figure}
\vspace{5mm}
 \hspace{-5mm} $(a)$  \hspace{36mm}   $(b)$\\
  \includegraphics[width=0.49\columnwidth]{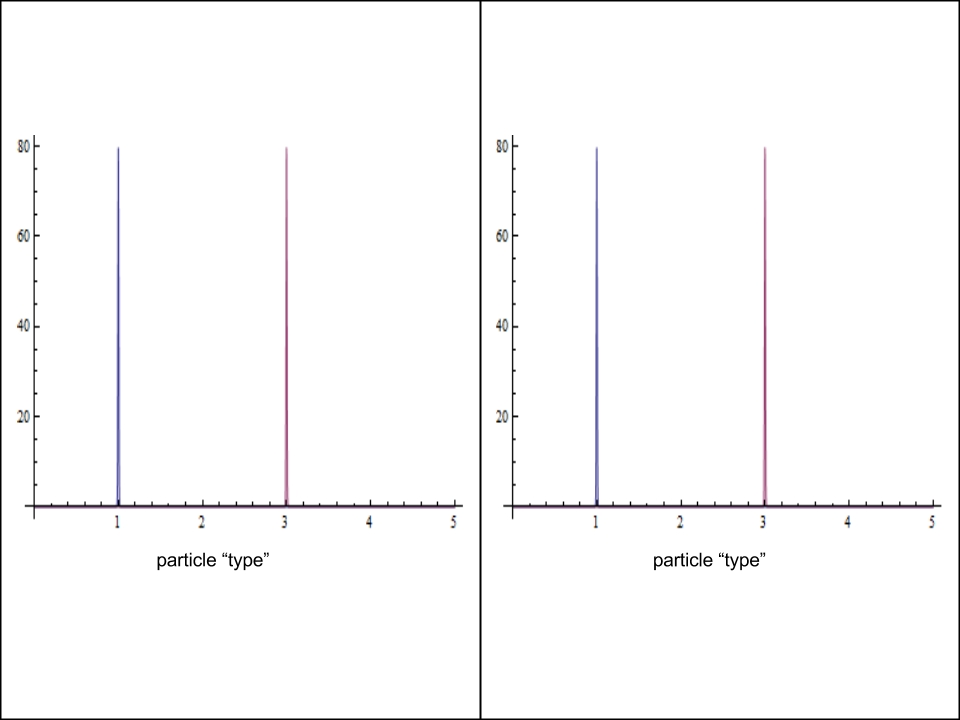}
  \includegraphics[width=0.49\columnwidth]{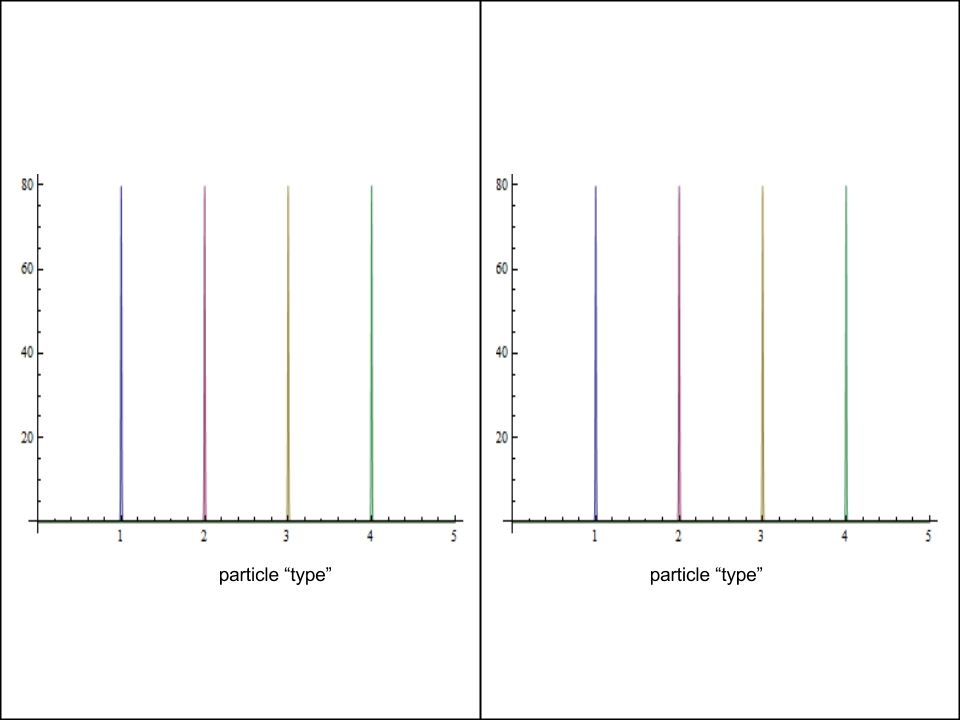}\\
 \hspace{-5mm} $(c)$  \hspace{36mm}   $(d)$\\ 
  \includegraphics[width=0.49\columnwidth]{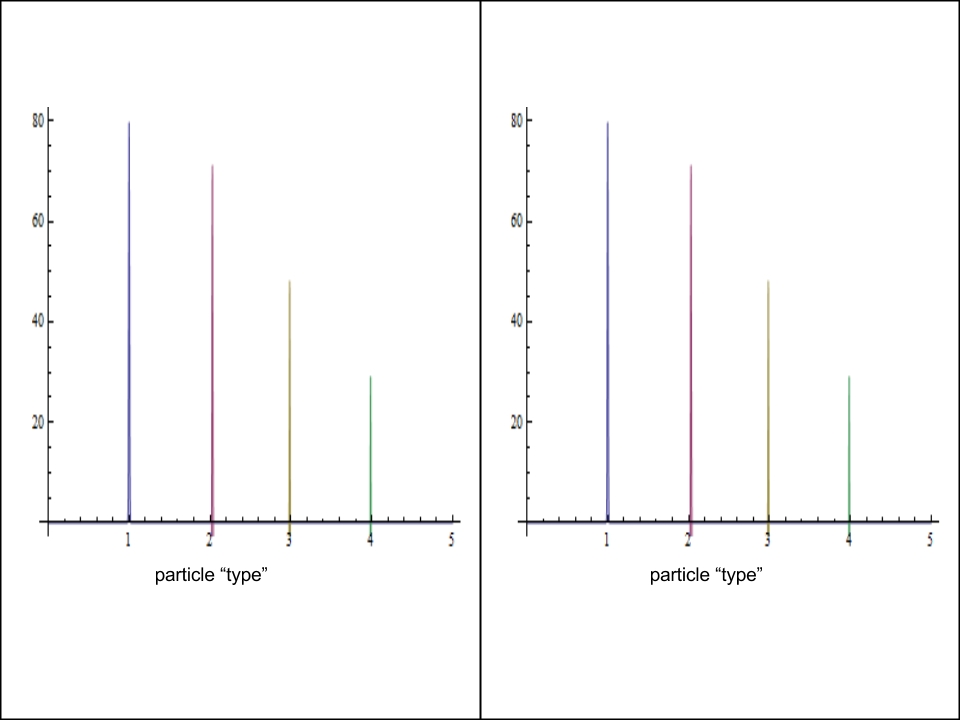}
   \includegraphics[width=0.49\columnwidth]{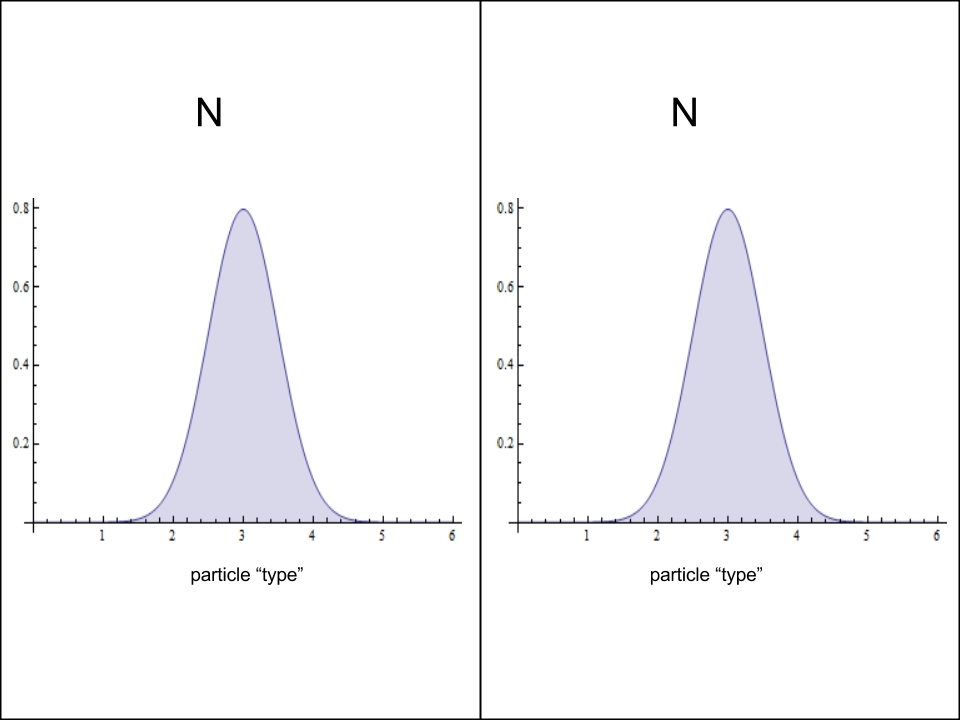}\\
  \caption{\label{fig2} Going from a {\it particle representation} to a {\it composition representation} of various realizations of case 2) of Gibbs' thought experiment. On the x-axis is schematically represented the particle {\it type} .}
\end{figure}
In Fig.\ref{fig2}, this last point is extended even further by looking at different cases where the assumption of zero mixing entropy should hold when two gases are identical in this broader sense. To do this, we schematically represent the state of the system with regards to the identity of its particles by the {\it composition} of the system expressed as a probability distribution of a random variable characterizing the {\it type} of the particle. This random variable can be anything ranging from the size of the particle to the number of neutrons it contains for instance. The three first compositions (a), (b) and (c) of Fig.\ref{fig2} have peaked distributions in their composition and correspond to discrete mixtures. The last plot (d), instead, corresponds to a continuous polydisperse system. But if  case d) corresponds to zero entropy change and it can be described --- for simplicity ---- as an ideal gas, then the only way to get the calculation right is by subtracting $\ln N!$ from its thermodynamic potential however polydisperse it may appear! Can we justify this? To answer this question, let us consider the following example: consider the mixing of a fully polydisperse gas (like one with composition described by (d) in Fig.\ref{fig2}) on the one hand, with a single component gas such that this component is very different from all the other particles in the polydisperse system \footnote{To imagine how this could be possible since the composition is continuous for the polydisperse system, it is enough to pick a single component system whose chemical type is not parametrized by the same random variable but by another one for which the polydisperse system has a peaked distribution.}. Mixing these two gases will inevitably lead to a non zero mixing entropy. Thermodynamically speaking, this {\it mixing entropy} corresponds to the work needed to separate back these two gases as they were {\it before} the mixing, if the gases are dilute enough \footnote{If the gases are not in a dilute phase, then the work provided will be equal to the free energy change instead of being simply proportional to the mixing entropy. In that case, one needs to determine first the excess free energy difference and only then it is possible to get the mixing entropy.}. In this case, we know the result from textbooks in the ideal gas case and $\Delta S_{mixing} = N\ln 2$. This factor $\ln 2$ in the entropy per particle that arises in statistical mechanics if we imagine each gas as being a single component, but yet different from the one it will be mixing with. This suggests that, when mixing is involved, what matters is how much the two {\it gases} differ from one another and not how each particle in these gases differ from {\it all} the others.

\section{The problem of the rational for N!}
As we have seen, from the epistemic point of view, the N! "fix" to the Gibbs' paradox seems to hold even for polydisperse systems since the notion of composition is itself a scale invariant property. As far as we know, this would be inconsistent with the quantum indistinguishability interpretation of the ``\ identical ''\ gases scenario. It is not clear however, how to predict what will be the mixing entropy in a more complicated case than the one depicted above leading to $\ln 2$. Also, lacking of a rational for the $\ln N!$ is not satisfactory if we want the statistical mechanics framework to be the procedure through which thermodynamics reduces to micromechanics. Let us first consider the partition function of a single component system at fixed $(N,V,\beta)$:
\begin{equation}
Q(N,\beta,V) \equiv \frac{V_{eff}(\beta,V)^N}{\mathcal{L}_d^{3N}} \equiv \frac{V^{N}}{\mathcal{L}_d^{3N}}  \int \frac{d^{3N}r}{V^N}\:e^{-\beta U(\bold{r_N})} \label{eq0}
\end{equation}where $\beta$ is the inverse temperature, $\mathcal{L}_d \equiv h/\sqrt{2 \pi m k_B T}$ the de Brooglie wavelength of the gas particles of mass $m$ (or any length characterizing the particle) and $U(\bold{r_N})$ the interaction potential between the particles in configuration $\bold{r_N}$.
If we assume that $V_{eff}(\beta,V)$ is extensive, then, upon scaling of the extensive variables by a factor $\lambda$, the free energy $F \equiv -\ln Q( N,\beta,  V) $ becomes $F_{\lambda} = -\ln Q( \lambda N,\beta,  \lambda V) = \lambda F(N,\beta,V) + N\lambda \ln \lambda$. Had we had $-\ln N! \sim -N\ln N + N$ in our expression of the free energy, it would have been extensive. Hence the ``\ right ''\ partition function from the point of view of thermodynamics seems to be $Q_{phys}(N,\beta,V)\equiv Q(N,\beta,V)/N!$. This can be understood by re-interpreting the microstates to be counted in the partition function as being solely non superimposable and non permutation-inversion isomer configurations in phase space \cite{Wales}.  
One can then easily see that the partition function of a mixture with $m$ species with the same weight in the composition reads:
\begin{equation}
Q_{phys}(N,\beta,V) \equiv \prod_{\alpha = 1}^m \left(\frac{V^{N/m}}{\mathcal{L}_{\alpha}^3 (N/m)!} \right) \int \frac{d^{3N}r}{V^N}\:e^{-\beta U(\bold{r_N})} \label{eq0bis}
\end{equation}where $\mathcal{L}_{\alpha}$ is the de Broglie wavelength of species $\alpha$ (or any length characterizing the particle of type $\alpha$). Note that all of this is very well known to anyone having followed any undergraduate course on statistical mechanics \cite{Huang, Balian, McQuarrie}. However, what may differ is the reasoning used to derive Eq. \eqref{eq0bis}. Although it corresponds to a standard result, it is usually presented as deriving {\it a priori} from an assumed more profound quantum statistical mechanics. Here, on the contrary, we derive it as the simplest form for the partition function that makes the free energy extensive. Hence, in a spirit following that of Khinchin for instance \cite{Khinchin}, the statistical mechanics framework is tested and refined from what we know about thermodyncamics and not the other way around. That is because there is, in principle, no objective way to decide what is a good or bad microstate to begin with.
Going back to Eq. \eqref{eq0bis}, the contribution of these $(N/m)!$ to the free energy is the following:
\begin{equation}
\ln \left[(N/m)!\right]^m \simeq N\ln N - N - N \ln m \label{eq1}  
\end{equation}where we used Stirling's approximation. Putting it back into Eq.\eqref{eq0bis}, one finds for the free energy:
\begin{equation}
 -\ln  Q_{phys} \simeq -\ln Q(N,\beta,V) + \ln N! -N \ln m \label{eq2}
\end{equation}It is interesting to note that we can retrieve the $N!$ in the large $N$ limit even if we are dealing with a mixture. The correction from the sole $\ln N!$ is $N\ln m$ and does not affect any of the thermodynamic quantities of the system at the exception of the chemical potential. Nevertheless, while the excess chemical potential of a mixture can be tricky to define \cite{Warren,Sollich01}, the combinatorial part we are discussing here is non ambiguous. In fact, since the combinatorial term $N\ln m$ only adds a constant to the chemical potential, it can be forgotten altogether since the chemical potential is defined up to a constant (at fixed composition). This allows then to rationalize more rigorously the bold arguments made in the first section where a $\ln N!$ would be subtracted from any thermodynamic potential characterizing one particular gas --- albeit polydisperse --- that could be discriminated surely from another gas with which it would mix experimentally.

 In the general case of a discrete mixture with $m$ species but unequal weights $\{p_{\alpha} \}_{\alpha=1..m}$ in the composition denoted $\mathcal{C}$, one can easily find that:
\begin{equation}
 -\ln  Q_{phys} \simeq -\ln Q(N,\beta,V) + \ln N! -Ns(\mathcal{C})\label{eq2}
\end{equation}where
\begin{equation}
s(\mathcal{C})\equiv -\sum_{\alpha}p_{\mathcal{C}}(\alpha)\ln p_{\mathcal{C}}(\alpha) \label{eq2bis}
\end{equation}has the shape of a Shannon entropy. We shall point out here that the expression \eqref{eq2bis} is sometimes called {\it mixing entropy}, name that we haven't kept in the present work as the composition under study does not need to be obtained by actual mixing; hence, we refer to it as {\it composition entropy}. We shall see in the next section that while the two concepts are related, they are not equivalent. 
Now, we can see that $s(\mathcal{C})$ is still a constant setting the reference of the chemical potential and should not play any role when looking at the thermodynamics of the system provided the composition remains unchanged.

\hspace{2mm}

Let us now turn to the subtle case of a continuous mixture where the previous interpretation, based on considering a subset of microstates that can't be related by particle permutations, cannot hold exactly and could easily lead to the conclusion that there shouldn't be any $N!$ arising in the problem ( we would then be back to what seems to be prescribed by the quantum interpretation). Yet, we shall claim here the contrary.

As a matter of fact, one point of discord is that it is mathematically impossible (with probability one) to have two particles that are {\it exactly} the same in a continuous polydisperse system, even in the thermodynamic limit \cite{Salacuse84}. Although some arguments can be made regarding the order of limits between the number of species and the number of particles \cite{Salacuse84, Sollich01}, this implies, in the worst case scenario, that it is not possible to generate an exact equivalence relation between states that are related by an element of the permutation group because there is none. Although this seems quite dramatic at first sight, this issue is far from new and is at the root of continuous probability theory where the probability for a continuous random variable to take a single value is always exactly zero. The solution traditionally used in mathematics consists (very roughly) in saying that although a continuous random variable will never take twice the same value, it will be as close to any value as anyone wants and there are thus many particles whose type lies say in the interval $[\alpha,\alpha+\Delta_{\alpha}]$ for instance. Let us denote this number $Nf_{\mathcal{C}}(\alpha)\Delta_{\alpha}$, where $f_{\mathcal{C}}(\alpha)$ is the probability density that characterizes the polydispersity, and assume it is much bigger than unity, then we can use results from the general discrete mixture case by replacing $p_{\mathcal{C}}(\alpha)$ with $f_{\mathcal{C}}(\alpha)\Delta_{\alpha}$ and find:
\begin{equation}
-\ln  Q_{phys} \simeq -\ln Q(N,\beta,V) + \ln N! -Ns(\mathcal{C},\Delta_{\alpha}) \label{eq3}
\end{equation}Physically speaking, $\Delta_{\alpha}$ represents the precision on the measurement of the ``\ type variable ''\ $\alpha$ and therefore the uncertainty to decide whether two microstates can be related by permuting a subset of the $N$ particles. This allows to weaken the criterion on which was built the equivalence relation before where now two configurations are considered equivalent if they can be related by swapping two particles whose $\alpha$ values are no further apart than $\Delta_{\alpha}$. Because of the continuous nature of the polydispersity, it is assumed that $\Delta_{\alpha} \ll |\alpha_{max}-\alpha_{min}|$ and therefore, the entropy sum in Eq.\eqref{eq3} can be very well approximated by an integral. In all generality, $\Delta_{\alpha}$ does not need to be uniform and the corresponding limiting integral reads:
\begin{equation}
s_{\Lambda}(\mathcal{C}) = -\int_{\alpha_{min}}^{\alpha_{max}} d\alpha \:f_{\mathcal{C}}(\alpha) \ln \Lambda(\alpha)f_{\mathcal{C}}(\alpha) \:\:\:\:(+\ln \mathcal{N}) \label{eq4}
\end{equation}where $\mathcal{N}$ and $\Lambda(\alpha)$ are such that $\Delta_{\alpha} \equiv \Lambda(\alpha)/\mathcal{N}$ with Eq. \eqref{eq4} obtained in the limit where $\mathcal{N}$ tends to infinity while $\Lambda(\alpha)$ keeps finite values. The general continuous entropy $s_{\Lambda}$ is thought to be originally due to Jaynes \cite{Jaynes63} and is sometimes referred to as the {\it Shannon-Jaynes} entropy. As we will see in the last section, the role of the --- unknown --- function $\Lambda$ can be crucial for practical purposes. It essentially refers to a prior assumption one makes about the ``\ denseness ''\ of $\alpha$-points in each interval $\Delta_{\alpha}$. In addition, although this prior knowledge can sometimes be modeled as a prior distribution, this is not a requirement to get reliable results \cite{Trizac11}. 

Finally, akin to what we have done  with discrete mixtures, we now note that the continuous composition entropy still does not affect the thermodynamic predictions one can make on the system as long as the composition is left unchanged and therefore we can forget about it and retrieve the simple $N!$ motivated at the beginning of the paper. This general point has also been stressed in past studies on phase equilibria in polydisperse systems where Eq. \eqref{eq4} is used as a starting point of the analysis \cite{Sollich98, Sollich01, Sollich02}.

In what follows, to avoid the use of too many notations, we shall use the same notation $p_{\mathcal{C}}(\alpha)$ to characterize either the discrete probability for the composition $\mathcal{C}$ or the corresponding probability density if $\alpha$ takes continuous values. 

\section{From composition entropies to mixing entropy}
In the previous section, we have seen that for any mixture (even continuous) at equilibrium, requiring extensivity of the thermodynamic potentials can always be done by subtracting a $\ln N!$ from the thermodynamic potentials and up to an additional constant corresponding to the composition entropy. We have also recognised that this additional constant could be discarded if one is only interested in those phases of the system where the composition $\mathcal{C}$ is unchanged. In this section, we instead look at the case of mixing where the composition changes and where we cannot disregard the combinatorial term characterizing the mixture.

Let us consider two gases with the same values of their scale invariant variables but different compositions respectively $\mathcal{C}_1$, with $N_1$ particles, and $\mathcal{C}_2$, with $N_2$ particles, that mix to yield a new composition $\mathcal{C}_3$.
We shall focus here on the combinatorial aspect of this mixing. Their thermodynamic entropies contain composition entropies (given by discrete or a continuous Shannon entropies) that we can not discard anymore and that read formally $N_1s_{\Lambda}(\mathcal{C}_1)$ and $N_2s_{\Lambda}(\mathcal{C}_2)$ respectively. Also, the final composition $\mathcal{C}_3$ has a new composition entropy $N s_{\Lambda}(\mathcal{C}_3)$. Looking solely at the combinatorial contribution to the mixing entropy, it is straightforward to see that the mixing entropy per particle is $s_{\Lambda}(\mathcal{C}_3)-(N_1s_{\Lambda}(\mathcal{C}_1)+ N_2s_{\Lambda}(\mathcal{C}_2))/N$. The composition $\mathcal{C}_3$ can in principle be anything. However, since $\mathcal{C}_3$ originates from mixing of $\mathcal{C}_1$ and $\mathcal{C}_2$, it is in fact related to them in some way. In absence of chemical reaction, the number of particle of each species is conserved and yields $p_{\mathcal{C}_3}(\alpha) = (N_1p_{\mathcal{C}_1}(\alpha)+N_2p_{\mathcal{C}_2}(\alpha))/N$. The total mixing entropy becomes then:
\begin{equation}
\Delta S_{mix}^{comb} = N_1 D_{KL}(\mathcal{C}_1||\mathcal{C}_3) + N_2 D_{KL}(\mathcal{C}_2||\mathcal{C}_3) \label{eq5}
\end{equation}where $D_{KL}(\mathcal{C}||\mathcal{C}')$ is the {\it Kullback-Leibler (KL) divergence} of composition $\mathcal{C}'$ from $\mathcal{C}$ sometimes referred as {\it relative entropy} and defined by \cite{KullbackLeibler}:
\begin{equation}
D_{KL}(\mathcal{C}||\mathcal{C}') \equiv -\sum_{\alpha} p_{\mathcal{C}}(\alpha) \ln \frac{p_{\mathcal{C}'}(\alpha)}{p_{\mathcal{C}}(\alpha)} \label{eq6}
\end{equation}Note that although the KL divergence is defined in Eq. \eqref{eq6} for discrete mixtures, its definition still holds for continuous ones and is {\it independent} of $\Lambda$ in that case.
Besides estimating how much the original compositions differ from the final one, the mixing entropy in \eqref{eq5} also accounts for the different weights each composition has in the mixing process. If we want to understand more deeply the role of the compositions, it is then sensible to choose an ``\ even ''\ mixing where $N_1 = N_2 = N/2$. In this case, Eq. \eqref{eq5} yields:
\begin{eqnarray}
\Delta S_{mix}^{comb} &=& N D_{JS}(\mathcal{C}_1||\mathcal{C}_2) \nonumber \\ 
&\equiv &  \frac{N}{2}\sum_{\alpha}\left[p_{\mathcal{C}_1}(\alpha)\ln \frac{2p_{\mathcal{C}_1}(\alpha)}{p_{\mathcal{C}_1}(\alpha) + p_{\mathcal{C}_2}(\alpha)} \right. \nonumber \\ && \left. +\: p_{\mathcal{C}_2}(\alpha)\ln \frac{2p_{\mathcal{C}_2}(\alpha)}{p_{\mathcal{C}_1}(\alpha) + p_{\mathcal{C}_2}(\alpha)} \right] \label{eq7}
\end{eqnarray}which defines $D_{JS}(\mathcal{C}_1||\mathcal{C}_2)$, the {\it Jensen-Shannon (JS) divergence} \cite{JensenShannon}.

\hspace{2mm}

Eqs. \eqref{eq5} and \eqref{eq7} constitute the principal result of the paper that allows one to express the combinatorial contribution to the mixing entropy in absence of chemical reactions in any possible case. We note that expression \eqref{eq5} was already derived in Ref. \cite{Sollich01} (see Eq. 36 of the reference) while the recasting of Eqs. \eqref{eq5} and \eqref{eq7} in terms of the KL and JS divergences is essentially new, to our knowledge, and will enable us to apprehend existing results in the fields of polydisperse systems and information theory in a new way as we shall demonstrate below.\newline One last comment remains in order: as we understand it, the equation akin to Eq. \eqref{eq5} in Ref. \cite{Sollich01} expresses, in the language of the present paper, the composition entropy \eqref{eq4} of a polydisperse system, comprising two coexisting compositions $1$ and $2$, when the function $\Lambda$ is chosen to be the inverse probability density of a parent composition labeled $0$. The whole problem is then naturally interpreted as a statistical inference problem where the ``\ right ''\ choice of prior is the parent composition. In our case however, Eq. \eqref{eq5} refers to a quantity, namely the mixing entropy, that {\it is not} in general the same as the composition entropy defined in Eq. \eqref{eq4} and is in fact independent of the choice of $\Lambda$. This distinction can become important when looking at cases, like chemical reactions, where the choice of $\Lambda$ really matters to make meaningful predictions on the system (cf. the last section of the present paper).


 \section{Highly asymmetric mixing}
A particular case of interest is that of mixing when one composition, say $\mathcal{C}_1$, is overwhelmingly more represented than the other $\mathcal{C}_2$ {\it i.e.} $N_1 \gg N_2$. As a consequence, the explicit ``\ symmetry ''\ in the roles played by $\mathcal{C}_1$ and $\mathcal{C}_2$  in Eq. \eqref{eq5} will disappear. To see this, it is more convenient to fix the total number of particles $N$ and rename $\epsilon = N_2/N$ and $N_1/N = 1-\epsilon$. The entropy difference between a mixed state $\mathcal{C}_3$ and the separated states reads then from Eq. \eqref{eq5}:
  \begin{equation}
 \Delta S_{mix}^{comb}(\epsilon) = N\epsilon D_{KL}(\mathcal{C}_2 || \mathcal{C}_3) + N (1-\epsilon)D_{KL}(\mathcal{C}_1||\mathcal{C}_3) \label{demix1}
 \end{equation}Now, the entropy change as expressed in Eq. \eqref{demix1} is vanishingly small since the change in the system is minute. In particular, it is easy to show that $D_{KL}(\mathcal{C}_2 || \mathcal{C}_3) \sim \mathcal{O}(1)$ and $D_{KL}(\mathcal{C}_1 || \mathcal{C}_3) \sim \mathcal{O}(\epsilon^2)$. The entropy change per particle upon mixing is then dominated by the entropy change of the dominated composition $\mathcal{C}_2$ such that the limit of $ \Delta S_{mix}^{comb}(\epsilon)/(N\epsilon)$ converges to:
  \begin{equation}
 \lim_{\epsilon \rightarrow 0} \frac{ \Delta S_{mix}^{comb}(\epsilon)}{N\epsilon} = D_{KL}(\mathcal{C}_2||\mathcal{C}_3), \label{demix2}
\end{equation}which is a very general and exact result for totally asymmetric mixing. It is worth noting that, from a statistical inference perspective, it is tempting to interpret Eq.\eqref{demix2} as the KL divergence of composition $\mathcal{C}_2$ from composition $\mathcal{C}_3$, but there is actually {\it a priori} no reason to consider one of these two as a reference more than the other since Eq. \eqref{demix2} comes from the general mixing entropy variation Eq. \eqref{eq5}. 

\vspace{2mm}
Let us now turn to a concrete application where Eq. \eqref{demix2} can be of use. Demixing phase transitions in polydisperse mixtures are usually characterized by the determination of the shadow and cloud curves. In these curves a majority, reference phase, coexists with an incipient phase. This thermodynamic equilibrium corresponds to a highly asymmetric demixing situation. The entropy change associated to bringing the majority (or parent) phase in coexistence with the incipient one is characterized by Eq.\eqref{demix2}, where $\mathcal{C}_3$ corresponds to the parent phase, $\mathcal{C}_1$ to the majority phase coexisting with $\mathcal{C}_2$, the incipient thermodynamic phase. In fact, this result has already been shown to be exact in the context of thermodynamic equilibria in polydisperse systems \cite{Sollich98, Sollich01, Sollich02}. In the present context, this expression for the entropy of mixing is derived as a by-product of the general strategy presented in the paper and shows its potential in different contexts where the combinatorial entropy is relevant.

\section{Link with the Landauer bound}
The combinatorial part of the mixing entropy expressed as a JS divergence \eqref{eq7}, for a mixing process involving equal proportions of the original compositions, has interesting properties. First, $\sqrt{D_{JS}(\mathcal{C}_1||\mathcal{C}_2)}$ is an actual {\it metric} in the space of probability densities \cite{JSmetric}. It implies that, as we argued in the first sections, what matters when mixing two gases is not the absolute identity of the particles involved in the mixing process but how the compositions of the two gases differ from one another; and this difference is in fact measured by the (combinatorial) mixing entropy per particle which corresponds to the square of a geometric distance between the two initial compositions. Secondly, for any two probability distributions $0 \leqslant D_{JS} \leqslant \ln 2$ \cite{JensenShannon} which means that the maximum contribution of permutations to the mixing entropy is $N k_B \ln 2$. In fact, in the particular case where the two compositions are genuinely distinct, $k_B T \ln 2$ represents the {\it minimum} (in absence of friction) amount of work to provide in order to separate the two mixed species into equally sized compartments (going from (c) to (a) in Fig. \ref{fig1}); which is reminiscent of Landauer's principle \cite{Landauer} of the minimum amount of energy to dissipate in order to erase one bit of memory.

We therefore seek an interpretation of Eq. \eqref{eq7} as the minimum energy dissipated to erase the information contained in the mixed state of two polydisperse gases. Such an interpretation has been made possible by a recent experimental work \cite{Landauer12} that showed that Landauer's bound could be approached very closely by very small and controlled systems and that the energy required to erase one bit of memory was, depending on the protocol, smaller than $k_B T \ln 2$. There is no violation of the Landauer's principle, however, since those protocols that are able to erase one bit of memory with less than $k_B T \ln 2$ energy do not succeed all the time. Hence what then matters is the {\it rate of success} of the erasure operation and the closer it is to unity, the closer the average work needed will be to $\ln 2$ \cite{Landauer12}. In our case of ideal particles in a box, we can construct a 1-bit memory unit for each particle by assigning a value zero or one to the states $x < L/2$ and $x \geqslant L/2$ respectively where $x$ and $L$ denote the $x$-coordinate and the box size along that direction. Since the gas is ideal, we can focus on each particle separately. At equilibrium in the mixed state, each particle is in a state represented by the density operator $\hat{\rho}_{mixed} = 1/2 |0\rangle \langle 0| + 1/2 |1\rangle \langle 1|$ where we used quantum mechanics notations only to make the reasoning more readable. The entropy of such a mixed state is $s_{mixed} \equiv - {\rm Tr}(\hat{\rho}\ln \hat{\rho}) = \ln 2 $. We now consider the case of a protocol whose goal is to erase the memory of this initial mixed state at the profit of another final state. One such protocol could for instance bring certainly the particle (whatever its type) from anywhere in the box to its right hand side. In that case, the corresponding final state is described by the density operator $\hat{\rho}_{1} = |1\rangle \langle 1|$ which has an entropy zero. Of course, the same specific protocol could be used to put certainly the system in a $|0\rangle$ state instead of a $|1\rangle$ state. The total change in free energy is $k_B T \ln 2$ and corresponds to the minimal work one has to provide to erase certainly the initial state which is the Landauer result. We now turn to a somewhat different protocol which has some uncertainty associated to it such that the probability for a particle of type $\alpha$ to end up on the right hand side of the box ({\it} i.e.) in state $|1\rangle$) is not exactly one. In fact, for a given particle with type $\alpha$, we are interested in the conditional probability that the particle will be in state $|1\rangle$ (or $|0\rangle$) once the protocol has ended and knowing that it is of type $\alpha$. To link this thought experiment with our mixing problem, let us consider the case of a demixing protocol where the probability to be of type $\alpha$ and on the right hand side of the box is $p_{\mathcal{C}_1}(\alpha)/2$ (and similarly, the probability to be of type $\alpha$ and on the left hand side of the box is $p_{\mathcal{C}_2}(\alpha)/2$) and that the probability to find a particle of type $\alpha$ is $p_{\alpha}(\mathcal{C}_3)=(p_{\mathcal{C}_1}(\alpha)+p_{\mathcal{C}_2}(\alpha))/2$ (in absence of chemical reaction). The corresponding conditional probability of success or {\it rate of success} \cite{Landauer12} is then  $p_{\mathcal{C}_1}(\alpha)/(p_{\mathcal{C}_1}(\alpha)+p_{\mathcal{C}_2}(\alpha))$. If the particle doesn't end up on the right hand side of the box, then it has to be on its left hand side with probability $1-p_{\mathcal{C}_1}(\alpha)/(p_{\mathcal{C}_1}(\alpha)+p_{\mathcal{C}_2}(\alpha)) = p_{\mathcal{C}_2}(\alpha)/(p_{\mathcal{C}_1}(\alpha)+p_{\mathcal{C}_2}(\alpha))$. The density operator corresponding to this final state is then: 

\begin{equation}
\hat{\rho}_{\alpha}(\mathcal{C}_1, \mathcal{C}_2) = \frac{p_{\mathcal{C}_2}(\alpha)}{p_{\mathcal{C}_1}(\alpha)+p_{\mathcal{C}_2}(\alpha)} |0\rangle \langle 0| + \frac{p_{\mathcal{C}_1}(\alpha)}{p_{\mathcal{C}_1}(\alpha)+p_{\mathcal{C}_2}(\alpha)} |1\rangle \langle 1| \label{erasure}
\end{equation}and the uncertainty it carries betrays the fact that some memory statistically remains from the original state. The corresponding entropy reads:

\begin{eqnarray}
s_{\alpha} (\mathcal{C}_1, \mathcal{C}_2) &=& -\frac{p_{\mathcal{C}_2}(\alpha)}{p_{\mathcal{C}_1}(\alpha)+p_{\mathcal{C}_2}(\alpha)} \ln \frac{p_{\mathcal{C}_2}(\alpha)}{p_{\mathcal{C}_1}(\alpha)+p_{\mathcal{C}_2}(\alpha)} \nonumber \\ 
&& - \frac{p_{\mathcal{C}_1}(\alpha)}{p_{\mathcal{C}_1}(\alpha)+p_{\mathcal{C}_2}(\alpha)} \ln \frac{p_{\mathcal{C}_1}(\alpha)}{p_{\mathcal{C}_1}(\alpha)+p_{\mathcal{C}_2}(\alpha)} \label{erasure1}
\end{eqnarray}For this protocol, Eq. \eqref{erasure1} holds for a particle knowing that it is of type $\alpha$. We now need to average the entropy in \eqref{erasure1} over the uncertainty related to the particle type (which is of a different nature than the probabilities associated to the final state which pertains to the erasure protocol). The corresponding expectation value for the entropy of the final state is then: $\langle s_{\alpha} (\mathcal{C}_1, \mathcal{C}_2) \rangle_{\mathcal{C}_3} \equiv \sum_{\alpha} p_{\alpha}(\mathcal{C}_3) s_{\alpha} (\mathcal{C}_1, \mathcal{C}_2)$ which gives for the work to provide in order to erase the initial state (in units of $k_B T$):

\begin{eqnarray}
\beta w(\mathcal{C}_1,\mathcal{C}_2) &\equiv & \ln 2 -\langle s_{\alpha} (\mathcal{C}_1, \mathcal{C}_2) \rangle_{\mathcal{C}_3}  \nonumber \\
&=& \ln 2 \nonumber \\
&& + \sum_{\alpha} \frac{p_{\mathcal{C}_1}(\alpha)}{2} \ln \frac{p_{\mathcal{C}_1}(\alpha)}{p_{\mathcal{C}_1}(\alpha)+p_{\mathcal{C}_2}(\alpha)} \nonumber \\
&& + \sum_{\alpha} \frac{p_{\mathcal{C}_2}(\alpha)}{2} \ln \frac{p_{\mathcal{C}_2}(\alpha)}{p_{\mathcal{C}_1}(\alpha)+p_{\mathcal{C}_2}(\alpha)} \label{erasure2}
\end{eqnarray}

Finally, rewriting $\ln 2$ as $1/2 \sum_{\alpha} p_{\mathcal{C}_1}(\alpha)\ln 2 + 1/2 \sum_{\alpha}  p_{\mathcal{C}_2}(\alpha)\ln 2 $ enables then to show that $\beta w(\mathcal{C}_1,\mathcal{C}_2) = D_{JS}(\mathcal{C}_1||\mathcal{C}_2)$. Hence, recent studies on the Landauer bound help us shedding light on how to interpret the appearance of the JS divergence in Eq. \eqref{eq7}. In particular, it can be understood as a deviation from the Landauer limit imposed on the energy dissipated to erase a 1-bit memory when the erasure protocol does not succeed with probability one. 

\vspace{2mm}

The presently drawn analogy between mixing (or demixing) and 1-bit memory manipulation enables us to argue that the mixing entropy of any two polydisperse systems --- albeit of finite size --- will follow on average the prescription given by Eq. \eqref{eq7}. In particular, it shows that it is impossible to extract useful and reproducible work from the mixing of any two finite realizations of the same underlying composition. This last point is important as it does not depend on a) whether or not we actually know the underlying distribution or b) whether or not the system is big enough to be a good representation of the distribution; rather, we simply need to know that the two realizations are generated by the same protocol (itself characterizable by a (known or unknown) probability distribution).

\section{Limiting behaviours of the JS divergence}
When looking at the entropy of an even mixing in the absence of chemical reactions, the combinatorial entropy change is proportional to the JS divergence between the two original compositions. This section aims at looking at different cases in which it can be relevant to estimate $D_{JS}$.
To warm up, we first consider the transformation from (a) to (c) in Fig.\ref{fig1}. There are only two values of $\alpha$ that we denote here as $\alpha_1$ and $\alpha_2$. The composition $\mathcal{C}_1$ on the left hand side is such that $(p_{\mathcal{C}_1}(\alpha_1)=1,p_{\mathcal{C}_1}(\alpha_1)=0 )$ and $\mathcal{C}_2$ on the right hand side is such that $(p_{\mathcal{C}_2}(\alpha_1)=0,p_{\mathcal{C}_2}(\alpha_2)=1 )$. We can now apply Eq.\eqref{eq7}:
\begin{equation}
D_{JS}(\mathcal{C}_1||\mathcal{C}_2) = \left(\frac{1}{2}\ln \frac{1}{1/2}+ \frac{1}{2}\ln \frac{1}{1/2}\right) = \ln 2 \label{eq8}
\end{equation}
which is the expected result. 

\hspace{2mm}

Let us now turn to the harder case of two continuous mixtures $\mathcal{C}_1$ and $\mathcal{C}_2$. We shall look at two different limits: when the two compositions are almost the same and when there is very little overlap between them.

\hspace{2mm}

When the compositions $\mathcal{C}_1$ and $\mathcal{C}_2$ are almost the same, it is easily shown that the JS divergence reads:
\begin{equation}
D_{JS}(\mathcal{C}_1||\mathcal{C}_2) = \left\langle \frac{h^2}{4} + \mathcal{O}(h^3) \right\rangle_{\overline{p}} \label{eq9}
\end{equation}where $\overline{p}(\alpha) \equiv (p_{\mathcal{C}_1}(\alpha)+p_{\mathcal{C}_2}(\alpha))/2 $, $h(\alpha) \equiv (p_{\mathcal{C}_1}(\alpha)-p_{\mathcal{C}_2}(\alpha))/\overline{p}$ and $\langle \cdot \rangle_{\overline{p}}$ is an average with respect to the probability density $\overline{p}$. Eq. \eqref{eq9} is quite insightful as we see that when two compositions are ``\ close ''\ to one another, the square of the mixing entropy per particle equates the mean of what is reminiscent of a Euclidean distance between the compositions. Incidentally, we also notice that when the two compositions are the same $h$ is identically zero and the JS divergence vanishes, as it should. 

\begin{figure}
  \includegraphics[width=0.95\columnwidth]{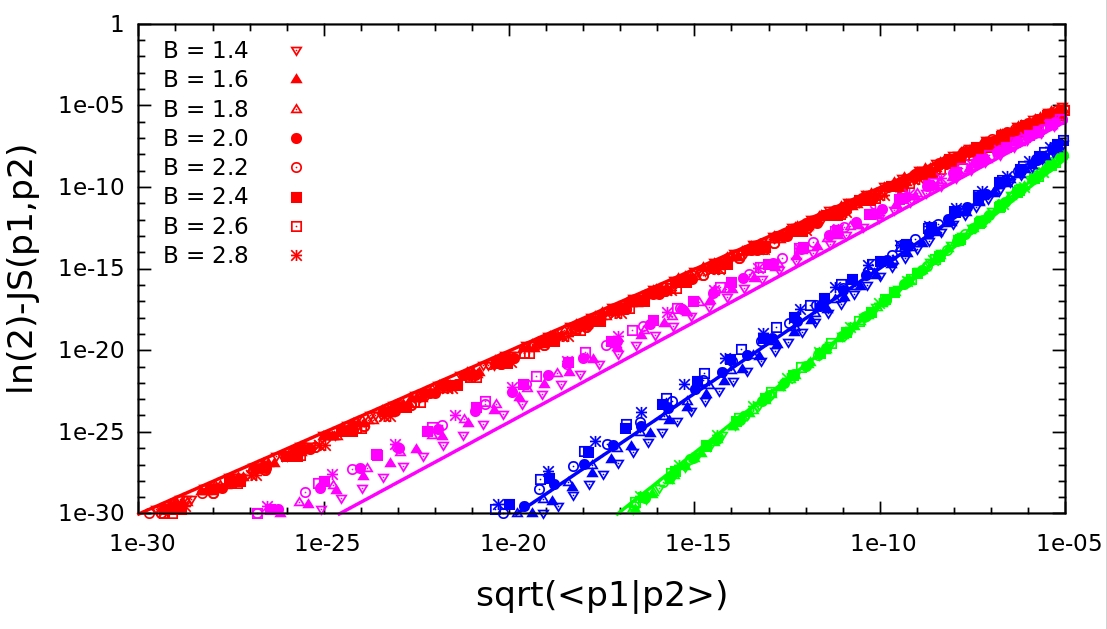} 
  \caption{\label{fig3} Plot in log-log scale of the $\ln 2 -D_{JS}$ divergence evaluated numerically as a function of $\sqrt{\langle p_{\mathcal{C}_1} | p_{\mathcal{C}_2} \rangle } = \sqrt{\langle \mu_1, \alpha_1, \beta | \mu_2, \alpha_2, \beta \rangle }$ also evaluated numerically in the case of generalized normal distributions for different values of the shape parameter $\beta$ and at fixed ratios $ r=\frac{\sigma_2}{\sigma_1}$ ($r = 1$ (red), $r=2$ (purple), $r=6$ (blue) and $r =27$ (green)). For a given ratio of the standard deviation $r_{12}$, all the $\beta$-values seem to fall on curves with an algebraic decay characterized by the same exponent very well approximated by Eq. \eqref{eqnu}. The corresponding curves are plotted in plain lines and compare relatively well with the numerical points.}
\end{figure}

\vspace{2mm}

Now, when the two compositions are almost distinguishable {e.g.} like the peaks of Fig. \ref{fig2}(a), then one has:
\begin{eqnarray}
\int d\alpha \:\left(p_{\mathcal{C}_1}+ p_{\mathcal{C}_2}\right)\ln \left( p_{\mathcal{C}_1}+p_{\mathcal{C}_2} \right) &\approx & \int d\alpha \:p_{\mathcal{C}_1}\ln p_{\mathcal{C}_1} \nonumber \\
&+&  \int d\alpha \:p_{\mathcal{C}_2}\ln p_{\mathcal{C}_2} \label{B0}
\end{eqnarray}which gives simply $\ln 2$ in Eq. \eqref{eq7}. More quantitatively, the fact that the compositions are ``\ far ''\ from one another can be characterized by a small overlap function that we denote $\langle f |g \rangle \equiv \int dx\:f(x)g(x)$ for any function $f$ and $g$. We performed a numerical analysis of the behaviour of the JS divergence as a function of the overlap function. To get some general insights on this behaviour, we chose the 3-parameter generalized normal distribution that reads:
\begin{equation}
GN(x|\mu,\alpha,\beta) \equiv \frac{\beta}{2\alpha \Gamma(1/\beta)}e^{-\left(|x-\mu |/\alpha \right)^{\beta}} \label{GN}
\end{equation}where $\Gamma(x)$ is the Euler Gamma function. The mean of this distribution is given by $\mu$, its variance by $\alpha^2 \Gamma(3/\beta)/\Gamma(1\beta)$ and the excess kurtosis by $\Gamma(5/\beta)\Gamma(1/\beta)/\Gamma(3/\beta)^2-3$. This familiy of symmetric distributions is interesting as it covers, among others, Laplace distributions when $\beta = 1$ and Normal distributions when $\beta = 2$. Fig. \ref{fig3} shows in particular that when looking at two composition probability distributions sharing the same values for $\beta$ but different ones for $\mu$ and $\alpha$, one would find a collapse of all the curves with the same ratio $r_{12} \equiv \sigma_2/\sigma_1$ on the same asymptotic algebraic decay for small overlaps:
\begin{equation}
\ln 2 - D_{JS}(\mathcal{C}_1||\mathcal{C}_2)  \sim \langle p_{\mathcal{C}_1}|p_{\mathcal{C}_2} \rangle^{\nu/2} \label{eq10}
\end{equation}where the exponent $\nu$ is empirically shown to satisfy:

\begin{equation}
\nu (r_{12}) \approx 1+ \tanh\left(\frac{1}{3}\ln r_{12} \right) \label{eqnu}
\end{equation}

Eqs. \eqref{eq9} and \eqref{eq10} thus provide limiting behaviours for the JS divergence tested in the case of generalized gaussian distributions that have the same value for the shape parameter $\beta$. The current analysis still lacks a proper understanding of Eq. \eqref{eq10} (and the subsequent more exploratory analysis of Appendix A) but it can prove useful to get more intuition about what the JS divergence depends upon in different regimes. In particular, it shows us that both the similarity in shape (characterized by Eq. \ref{eq9}) and the similarity in ``\ position ''\ in the space of the random variable (characterized by the overlap function), which are respectively dominant in one of the regime studied, contribute in an intricate way to the total JS divergence.

\section{The problem of chemical reactions}
We finally look at the case of chemical reactions between two compositions $\mathcal{C}_1$ and $\mathcal{C}_2$ which will mix and react by exchanging the property $\alpha$ according to some conservation law $C(\alpha)$. The total entropy variation for the $\{ \rm mixing + reaction \}$ is $\Delta S_{mix}^{comb} + \Delta S_{r}$ where $\Delta S_{mix}^{comb}$ is given by Eq. \eqref{eq5} and $\Delta S_{r}$ by:
\begin{equation}
\Delta S_{r} \equiv s_{\Lambda}(\mathcal{C}') -s_{\Lambda}(\mathcal{C}_3) \label{reaction0}
\end{equation}where $\mathcal{C}'$ is the new equilibrium composition once chemical equilibrium has been reached. We first notice that the type of chemical reactions considered here completely forgets about the initial composition apart from the quantity $C(\alpha)$ taking its value from it. If we want to find the equilibrium composition by maximizing the entropy variation, the situation becomes then equivalent to maximizing the composition entropy $s_{\Lambda}(\mathcal{C}')$ alone. This problem is at the root of a fundamental issue in statistical inference as the entropy to maximize depends on the unknown quantity $\Lambda$. Note that in the case of mixing without chemical reaction this would not have been an issue because the existence of an infinite number of conserved quantities (the number of particles per species) would allow us to write the sought new composition directly as a function of the two original --- known --- compositions $\mathcal{C}_1$ and $\mathcal{C}_2$ only; which makes the corresponding entropy variation explicitly independent of $\Lambda$ (cf. Eq. \eqref{eq5}). In general that is not the case and it is quite tricky to find what should be the best $\Lambda$ to use \cite{ Caticha04, Trizac11}. This pertains to the fact that the use of constraints in the optimization problem is not enough to characterize a unique composition $\mathcal{C}'$ \cite{Caticha04, Trizac11} and therefore one needs to rely on some {\it a priori} knowledge about the system, that is, propose a reasonable $\Lambda$ for the system under study. In Appendix B, we give a simple example to illustrate the difficulty of finding the right $\Lambda$ when dealing with problems unrelated to statistical thermodynamics.


\subsection{Equilibrium statistical mechanics as a prior}
In a very general case, finding an objective $\Lambda$ without further knowledge of the system is close to impossible. However, we are dealing with a physical system to which apply the rules of equilibrium statistical thermodynamics. In particular, the function $\Lambda^{-1}(\alpha)$ can be thought of as the {\it a priori} ``\ denseness ''\ of the points in the neighbourhood of the value $\alpha$. If $\alpha$ is a mesoscopic variable, then $\Lambda^{-1}(\alpha)$ corresponds to a weighted degeneracy function where the degeneracy is summed over the {\it microstates} of the system under study. Assuming that the fundamental principle of equiprobability holds for any microstate sharing the same $\alpha$ value, the ``\ best ''\ choice for $\Lambda^{-1}(\alpha)$ is likely to be density of state $\omega(\alpha)$.
One can then update the composition of the system knowing $\omega(\alpha)$ and the constraint $C(\alpha)$ by maximizing the functional:
\begin{equation}
\Delta \tilde{S}_{r} \equiv s_{\omega^{-1}}(\tilde{\mathcal{C}}') -s_{\omega^{-1}}(\mathcal{C}_3)-\lambda \int d\alpha \tilde{p}_{\mathcal{C}'}(\alpha)C(\alpha) \label{reaction1}
\end{equation}

\subsection{Posing the problem in a closed form}
A more pragmatical point of view, not limited to statistical thermodynamics, consists in noticing that the problem posed assumes the compositions $\mathcal{C}_1$ and $\mathcal{C}_2$ to be known. One can always wonder how their respective probability densities were found in the first place. In fact, in principle, their functional form has to either, be imposed experimentally or be inferred from combining data with a particular model whose functional form can be inferred from maximizing a $\Lambda$-entropy with certain constraints. We shall assume that in any case, there exists a set of constraints $\{C^1_i\}_{i=1..N_c}$ such that, $p_{\mathcal{C}_1}$ for instance, satisfies a maximum entropy principle and therefore reads \cite{Trizac11}:
\begin{equation}
p_{\mathcal{C}_1}(\alpha) = \Lambda^{-1}(\alpha)e^{-\sum_i \lambda^1_i C^1_i(\alpha)} \label{reaction2}
\end{equation}This allows to pose the problem of finding the new composition at chemical equilibrium in a closed form involving only known quantities:
\begin{eqnarray}
\Delta \tilde{S}_{r}& \equiv & D_{KL}(\mathcal{C}'||\mathcal{C}_1) - D_{KL}(\mathcal{C}_3||\mathcal{C}_1) \nonumber \\ && + \sum_i \lambda^1_i \int d\alpha (\tilde{p}_{\mathcal{C}'}-p_{\mathcal{C}_1}) - \lambda \int d\alpha \tilde{p}_{\mathcal{C}'} C(\alpha) \label{reaction3}
\end{eqnarray}This can then be maximized to give an answer consistent with the knowledge one has on the system from the start.

\vspace{2mm}

It is worth pointing out that the two proposed methods give the same results for the equilibrium composition and the entropy variation if the original compositions are inferred from equilibrium statistical mechanics by using the equal probability assumption, otherwise they may give different outcomes.

\section{Conclusion}
In this paper, we addressed the problem of the combinatorial entropy of polydisperse systems. Following an epistemic line of reasoning, we dismissed the claim (as many have done before us) --- based on the quantum indistinguishability of particles --- that consists in rejecting a zero entropy variation for a continuous polydisperse system mixing with a system of the same composition. This further leads us to prescribe that for the thermodynamic potentials of a system to be extensive, it was sufficient to simply subtract $\ln N!$ from them however polydisperse the system they characterize is. The theoretical ``\ correction ''\ to this prescription corresponds to the composition entropy that we introduce and that we do not assimilate to the mixing entropy. We however showed that it does not contribute to the thermodynamics of the system at fixed composition which is supported by the fact that entropies are defined up to an additional constant. In fact, we then showed that when the composition can change and all the species are conserved, the absolute composition entropies only appear in relative quantities involving KL divergences. That is because, one cannot choose two different entropies of reference when mixing two different compositions. We then focused on the case of even mixing which allowed us to show that the corresponding entropy variation was in fact a metric in the space of composition probability densities. We proposed then a mapping between the problem of mixing in thermodynamics and that of the erasure of a 1-bit memory when the erasure process is not always successful. In addition to being insightful, this mapping allowed us to stress that it is impossible to get any reproducible work out of the mixing of two finite size realizations of the same polydispersity.
 To complete the mixing analysis, we proposed limiting expressions of the JS metric in the case of general gaussian distributions that share the same $\beta$-parameter. These estimates can become handy in practical cases as these distributions together with the KL and JS divergences are used in many areas of applied mathematics and in particular in data analysis \cite{Runnalls07, Yogan05, Ofran06, ATLAS09}. Finally, we discussed the challenging problem of mixing with chemical reactions deleting most of the memory and proposed two consistent ways to address the problem at hand.

We believe this results to be important a) at a foundational level as they strengthen the already existing claims that the resolution of the Gibbs' paradox does not necessary lie in quantum indistinguishability and by stressing the crucial role of the prior assumption in maximizing an entropy, b) at a fundamental level as they propose a general formula for the mixing entropy of two polydisperse gases and give an interpretation of it in term of an extended Landauer's principle that can be used by the two communities involved and c) at a technical level by providing insightful limiting laws for the JS metric and potentially helpful approximate expressions for scientists involved in extensive data analysis.

\begin{acknowledgments}
{\small \it F. Paillusson acknowledges fruitful discussions with D. Frenkel, S. Martiniani and J-M. Victor. We acknowledge the Direccion General de Investigacion (Spain) and DURSI for financial support under
projects FIS 2011-22603 and 2009SGR-634, respectively.
 I.P. ackowledges financial support from Generalitat de Catalunya under program Icrea Acad\`emia.}
 \end{acknowledgments}
\section*{Appendix A}
\begin{figure}
  \includegraphics[width=0.95\columnwidth]{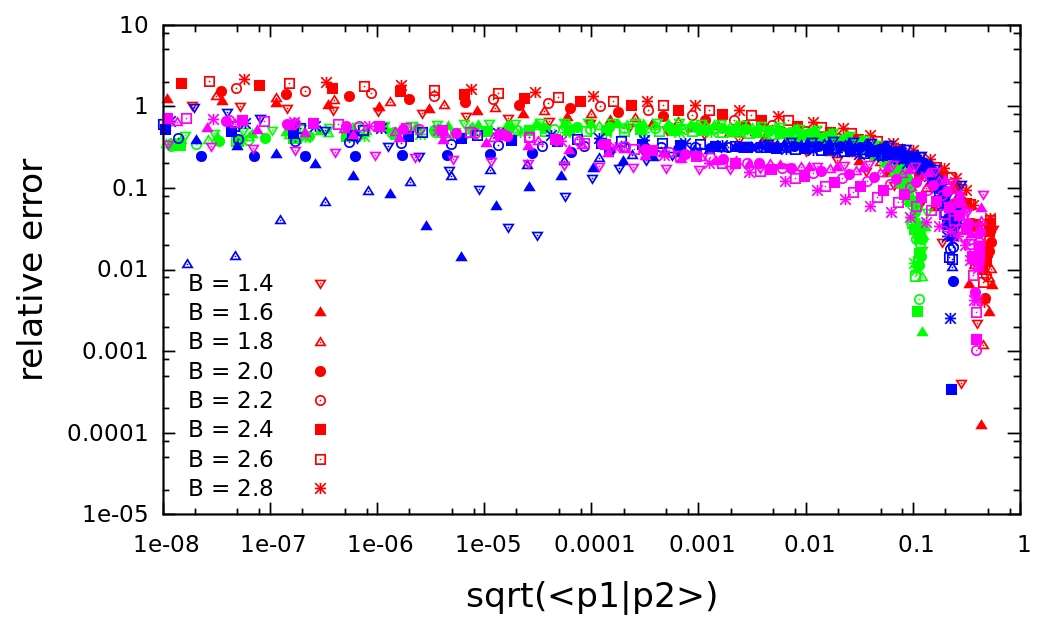} 
  \caption{\label{fig4} Plot of the relative error between the estimate in Eq. \eqref{eq11} and the numerically calculated JS divergence $|D_{JS}-D^{guess}_{JS}|/D_{JS}$ as a function of the overlap. The relative error is here shown for the same ratio values as that of Fig.\ref{fig3}.}
\end{figure}

This appendix aims at exploring the possibility of quantitative estimates of the JS divergence as a function of the overlap as the asymptotic behaviour of the former can be approximately described by \eqref{eq10} for various values of $\beta$ and $r_{12}$. It is tempting to search for a more general simple expression whose limit would be \eqref{eq10} for small overlaps and that would give a reasonable estimate of $D_{JS}$ for all ranges of interest.  
Fig. \ref{fig4} demonstrates that the following form is remarkably successful:
\begin{equation}
\tag{A1} D_{JS}(\mathcal{C}_1||\mathcal{C}_2) \approx \ln 2 - \left[ e^{(\varphi(r_{12})\sqrt{\sigma_{12}\langle p_{\mathcal{C}_1} |p_{\mathcal{C}_2} \rangle})^{\nu}}-1 \right] \label{eq11}
\end{equation}where $\sigma_{12} \equiv \sigma_1 + r_{12}\sigma_2$ and the function $\varphi$ has a value close to $\ln 2$ when $r_{12}$ is close to $1$ but decays as $r_{12}$ increases. The function $\varphi$ can be optimized to best fit either the asymptotic behaviour of the JS divergence or its values close to unity; this simple fact betrays the limitations of the hypothesis proposed in Eq. \eqref{eq11}.
\begin{figure}
  \includegraphics[width=0.95\columnwidth]{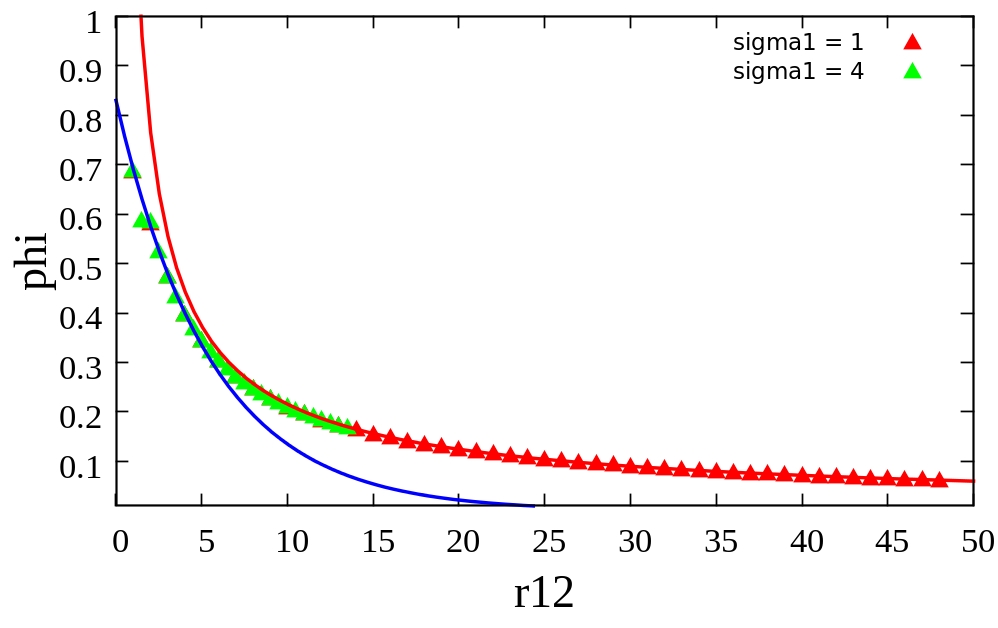} 
  \caption{\label{fig5} Plot of the function $\varphi$ as a function of $r_{12}$ for two different values of $\sigma_1 =1$ (red symbols) and $\sigma_1 = 4$ (green symbols). We see that the two curves $\varphi(\sigma_1 = 1, r_{12})$ and $\varphi(\sigma_1 = 4, r_{12})$ are superimposed which suggests that it is a function of $r_{12}$ only. The plain lines correspond to the best fits with the expression of Eqs. \eqref{eq12} (blue) and \eqref{eq13}.}
\end{figure}
However, from a pragmatic point of view, we chose to find the best $\varphi$ which is able to give reliable values of the JS divergence when it has values close to one and focus on the exponent when it has negligible values ($\sim$ $10^{-10}$). 
Figure \ref{fig5} shows that $\varphi$ decays first exponentially and then algebraically when $r_{12} > 5 $. It can be well approximated with the following empirical expression:
\begin{align}
\tag{A2} && \varphi(x) \approx 0.7 e^{-0.18x}, \hspace{2cm} x \leqslant 5 \label{eq12} \\
\tag{A3} &&\varphi(x) \approx 1.27 x^{-0.77}, \hspace{2cm} x > 5 \label{eq13} 
\end{align} It is worth pointing out that the function displayed in Fig. \ref{fig5} and the corresponding best fitting parameter values are independent of both $\beta$ and the individual values of $\sigma_1$ and $\sigma_2$ and shall provide, when used with Eq. \eqref{eq11}, a good estimate of the JS distance of generalized gaussian distributions that share the same $\beta$-value (as shown in Fig. \ref{fig4}). When allowing the parameter $\beta$ to vary between the two distributions, the phenomenology changes substantially with curves that use to belong to one master curve splitting into multiple branches with different slopes on a log-log scale. The actual approximate shape of Eq. \eqref{eq11} seems to remain valid and we could see that, regardless of the conditions, $1 \leq \nu \leq 2$.

\hspace{2mm}

We have thus proposed an empirical formula that is able to give reliable estimates of the JS divergence as a function of the overlap function between two GN distributions that share the same value for the parameter $\beta$. Considering the difficulty to get analytical estimates of the JS divergence even for gaussian distributions, we believe this result to be potentially useful for anybody needing a fast (albeit not exact) estimate of it. In particular, they can come handy in the field of Machine Learning and Data Analysis for classification and clustering problems where such estimates are sought \cite{Runnalls07} and which are becoming increasingly used in physical problems requiring extensive data analysis, especially in computational biophysics (see {\it e.g.} \cite{Yogan05, Ofran06} and high energy physics ({\it e.g.} \cite{ATLAS09}).

\section*{Appendix B}
The goal of this appendix is to show the difficulty that one encounters when using the Maximum Entropy Principle (MEP) for a distribution he knows {\it a priori} nothing about. We will use the paradigmatic example of the distribution of first digits of a sequence of uncorrelated random numbers taken from unrelated data sets (stock prices, number of inhabitants in countries, masses of objects from stars to molecules etc...).
Let us call $p(x)$ the probability density to have a real number with value $x$ in the range $R_x$ spanned by the numbers under investigation. We now call $d_1$ the first digit of $x$ in base $b$ such that $x = d_1 b^{m(x)} + r(x)$ where $m(x)$ is here the highest power of the base $b$ involved in the coding of $x$ and $r(x)$ is the rest of the digits to encode $x$. We define then, the probability $P(d_1|b)$ of having a number expressed in base $b$ to have $d_1$ as its first digit as:

\begin{equation}
\tag{B1} P(d_1|b) = \lim_{\epsilon \rightarrow 0} \sum_{m=m_{min}}^{m_{max}} \int_0^{b^m - \epsilon} dr \:p(d_1 b^m + r)  \label{B1}
\end{equation}

We now try to figure out $p(x)$ by maximizing its {\it Shannon-Jaynes} entropy $s_{\Lambda}$. Since there is no constraint on the numbers under study, it is reasonable to use Pascal's ``\ Principle of indifference ''\ and use a uniform prior $\Lambda(x) \equiv 1$. Together with the constraint of being normalized, the MEP yields $p(x) = R_x^{-1}$ which gives rise to a uniform --- yet unknown --- conditional probability $P(d_1|b)$. Using then the fact that the latter has to be normalized as well yields $P(d_1|b) = (b-1)^{-1}$. This uniform distribution for the first digit of numbers taken from many different sources sounds reasonable although it is most of the time wrong, as first noticed by Benford \cite{Benford38, Genome12}. 

In fact, a better way to look at the problem consists in noticing that since the numbers are very different and span many orders of magnitudes, their frequency of occurrence is more reliably described if one assumes a uniform prior for the variable characterizing the order of magnitude of their value {\it i.e.} a uniform distribution in logarithmic scale. It is straightforward to see that this is equivalent to considering a prior inversely proportional to $x$ {\it i.e.} $\Lambda(x) = x$. By applying the MEP, we then find (assuming $x \neq 0$) $p(x) = K/x$ where $K$ is a normalization constant.

Putting it back into Eq. \eqref{B1}, we get:

\begin{equation}
\tag{B2} P(d_1|b) = K' \ln \left(1 + \frac{1}{d_1} \right) \label{B2}
\end{equation}where $K'$ is a normalization constant that contains $K$. Finally requiring normalization of $P(d_1|b)$ yields:

\begin{equation}
\tag{B3} K'\sum_{n = 1}^{b-1} \ln \left(1 + \frac{1}{n} \right) = K'\sum_{n = 1}^{b-1} \ln \left(\frac{1+n}{n} \right) = K' \ln b = 1 \label{B3}
\end{equation}which allows us to retrieve the Benford's law for the probability of having the first digit as $d_1$ for a set of very different numbers expressed in basis $b$:

\begin{equation}
\tag{B4} P(d_1|b) = \ln_b \left(1 + \frac{1}{d_1} \right). \label{B4}
\end{equation}

With this simple example outside of the realm of equilibrium statistical mechanics, we have thus illustrated the care that needs to be used in choosing the quantity $\Lambda$ so as to infer the right probability density for the system under investigation.

\vspace{5mm}
 
\bibliographystyle{unsrt}
\bibliography{biblio_granular}

\begin{thebibliography}{10}

\bibitem{Salacuse84}
J.~J. Salacuse.
\newblock Random systems of particles: An approach to polydisperse systems.
\newblock {\em J.Chem.Phys.}, 81:2468, 1984.

\bibitem{Sollich98}
P.~Sollich and M.~Cates.
\newblock Projected free energies for polydisperse phase equilibria.
\newblock {\em Phys. Rev. Lett.}, 80:1365, 1998.

\bibitem{Sollich01}
P.~Sollich, P.~B. Warren, and M.~Cates.
\newblock Moment free energies for polydisperse systems.
\newblock {\em Adv. Chem. Phys.}, 116:265, 2001.

\bibitem{Sollich02}
P.~Sollich.
\newblock Predicting phase equilibria in polydisperse systems.
\newblock {\em J. Phys. Cond. Matt.}, 14:79--117, 2002.

\bibitem{Warren}
P.~B. Warren.
\newblock {\em Phys. Rev. Lett.}, 80:1369, 1998.

\bibitem{Will}
W.~Jacobs and D.~Frenkel.
\newblock Predicting phase behavior in multicomponent mixtures.
\newblock {\em J. Chem. Phys.}, 139:024108, 2012.

\bibitem{Ignacio10}
M.~R. Evans, S.~N. Majumdar, I.~Pagonabarraga, and E.~Trizac.
\newblock Condensation transition in polydisperse hard rods.
\newblock {\em J.Chem.Phys.}, 132:014102, 2010.

\bibitem{Uhlenbeck32}
G.~E. Uhlenbeck and L.~Gropper.
\newblock The equation of state of a non ideal {Einstein-Bose} of {Fermi-Dirac}
  gas.
\newblock {\em Phys.Rev.}, 41:79, 1932.

\bibitem{Cohen-Tannoudji}
C.~Cohen-Tannoudji, B.~Diu, and F.~Lalo{\'e}.
\newblock {\em Quantum Mechanics}.
\newblock Wiley VCH, 2006.

\bibitem{Seglar14}
P.~Seglar and E.~P{\'e}rez.
\newblock Classical limit of the canonical partition function.
\newblock {\em Eur.J. Phys.}, 35:015004, 2014.

\bibitem{Balian}
R.~Balian.
\newblock {\em {From Microphysics to Macrophysics: Methods and applications of
  statistical physics}}.
\newblock Springer, 2003.

\bibitem{Huang}
Kerson Huang.
\newblock {\em {Statistical Mechanics}}.
\newblock John Wiley \& Sons, second edition, 1987.

\bibitem{Jaynes92}
E.~T. Jaynes.
\newblock The {Gibbs}' paradox.
\newblock In C.R. Smith, G.J. Erickson, and P.O. Neudorfer, editors, {\em
  Maximum Entropy and Bayesian Methods}, pages 1--22. Kluwer Academic, 1992.

\bibitem{Swendsen06}
R.~H. Swendsen.
\newblock {\em Am. J. Phys.}, 74:187--190, 2006.

\bibitem{vanKampen84}
N.~G. van Kampen.
\newblock The {Gibbs}' paradox.
\newblock In W.E. Parry, editor, {\em Essays in Theoretical Physics: in Honor
  of Dirk ter Haar}. Oxford: Pergamon, 1984.

\bibitem{Daan14}
D.~Frenkel.
\newblock Why colloidal systems can be described by statistical mechanics: Some
  not very original comments on the {Gibbs}' paradox.
\newblock {\em arXiv:1312.0206}.

\bibitem{McQuarrie}
D.~A. McQuarrie.
\newblock {\em {Statistical Mechanics}}.
\newblock University Science Books, first edition, 2000.

\bibitem{Wales}
D.~Wales.
\newblock {\em {Energy Landscapes: Applications to Clusters, Biomolecules and
  Glasses}}.
\newblock Cambridge University Press, 2004.

\bibitem{Khinchin}
A.~I. Khinchin.
\newblock {\em {Mathematical Fundations of Statistical Mechanics}}.
\newblock Dover Publications, 1949.

\bibitem{Jaynes63}
E.T Jaynes.
\newblock Information theory and statistical mechanics.
\newblock {\em Brandei University Summer Institute Lectures in Theoretical
  Physics}, 3:181, 1963.

\bibitem{Trizac11}
P.~Maynar and E.~Trizac.
\newblock Entropy of continuous mixtures and the measure problem.
\newblock {\em Phys. Rev. Lett.}, 106:160603, 2011.

\bibitem{KullbackLeibler}
S.~Kullback and R.A. Leibler.
\newblock On information and sufficiency.
\newblock {\em Ann. Math. Statist.}, 22:79--86, 1952.

\bibitem{JensenShannon}
J.~Lin.
\newblock Divergence measures based on the shannon entropy.
\newblock {\em Trans.Inf. Theory}, 37:145--151, 1991.

\bibitem{JSmetric}
D.~M. Endres and J.~E. Schindelin.
\newblock A new metric for probability distributions.
\newblock {\em Trans.Inf.Theory}, 49:1858--1860, 2003.

\bibitem{Landauer}
R.~Landauer.
\newblock Irreversibility and heat generation in the computing process.
\newblock {\em IBM Journal of Research and Development}, 5:183, 1961.

\bibitem{Landauer12}
A.~B{\'e}rut, A.~Arakelyan, A.~Petrosyan, S.~Ciliberto, R.~Dillenschneider, and
  E.~Lutz.
\newblock Experimental verification of {Landauer}'s principle linking
  information and thermodynamics.
\newblock {\em Nature}, 483:187, 2012.

\bibitem{Caticha04}
A.~Caticha and R.~Preuss.
\newblock Maximum entropy and bayesian data analysis: Entropic priors.
\newblock {\em Phys.Rev.E}, 70:046127, 2004.

\bibitem{Runnalls07}
A.R. Runnalls.
\newblock Kullback-{Leibler} approach to gaussian mixture reduction.
\newblock {\em IEEE Trans. Aero.Elec.Syst.}, 43:989--999, 2007.

\bibitem{Yogan05}
C-J. Ku and G.~Yona.
\newblock The distance-profile representation and its application to detection
  of distantly related protein families.
\newblock {\em BMC Bioinf.}, 6:282, 2005.

\bibitem{Ofran06}
Y.~Ofran and H.~Margalit.
\newblock Proteins of the same fold and unrelated sequences have similar amino
  acid composition.
\newblock {\em PROTEINS: Structure, function and Bioinformatics}, 64:275, 2006.

\bibitem{ATLAS09}
Software validation structure for the atlas high-level trigger.
\newblock {\em ATL-DAQ-PROC-2009-024, ATL-COM-DAQ-2008-018}, 2009.

\bibitem{Benford38}
F.~Benford.
\newblock The law of anomalous numbers.
\newblock {\em Proc.Am.Phil.Soc.}, 78:551, 1938.

\bibitem{Genome12}
J.L. Friar, T.~Goldman, and J.~P{\'e}rez-Mercader.
\newblock Genome sizes and the benford distribution.
\newblock {\em PLoS One}, 7:36624, 2012.

\end{thebibliography}

\end{document}